\documentclass{iopjournal}

\usepackage{ragged2e}
\usepackage{amssymb}
\usepackage{amsmath}
\usepackage{pgfplots}

\usepackage{comment}

\usepackage{tabularx}
\usepackage{longtable} % Tables that can span several pages
\usepackage{colortbl}
\usepackage{multirow}
\usepackage[table]{xcolor}
\usepackage{booktabs}
\definecolor{bluePoli}{cmyk}{0.4,0.1,0,0.4}

\usepackage{graphicx}
\usepackage{transparent} % Enables transparent images
\usepackage{eso-pic} % For the background picture on the title page
\usepackage{tikz} % A package for high-quality hand-made figures.
\usetikzlibrary{}
\graphicspath{{./Images/}} % Directory of the images
\usepackage{caption} % Coloured captions
\usepackage{subcaption}
\usepackage{xcolor} % Coloured captions
\usepackage{amsthm,thmtools,xcolor} % Coloured "Theorem"
\usepackage{float}
\usepackage{epigraph}

\usepackage{pdfpages} % To include a pdf file
\usepackage{afterpage}
\usepackage{lipsum} % DUMMY PACKAGE
\usepackage{fancyhdr} % For the headers
\usepackage{comment}
\usepackage{epstopdf}

\newcommand{\sep}{,\ }

\pgfplotsset{compat=1.18}
\begin{document}
\justifying
\articletype{Paper} %	 e.g. Paper, Letter, Topical Review...
%[TRIMEG-C1 GPU]
\title{OpenMP GPU Acceleration and Portability of TRIMEG-C1 for Electromagnetic Gyrokinetic Simulations in Tokamak Plasmas}

\author{Giorgio Daneri$^{1,*}$, Zhixin Lu$^{1,*}$, Matthias Hoelzl$^{1,2}$, Luca Venerando Greco$^1$, Edoardo Carr\`a$^1$}

\affil{$^1${Max Planck Institute for Plasma Physics, Boltzmannstr. 2,  Garching, 85748, Germany}}\\
\affil{$^2${Department of Physics and Astronomy, Chalmers University of Technology, Göteborg, SE-41296, Sweden}}

\affil{$^*$Author to whom any correspondence should be addressed (giorgio.daneri@ipp.mpg.de; zhixin.lu@ipp.mpg.de).}

%\email{name@institution.org}

\keywords{GPU acceleration\sep OpenMP offloading \sep GPU portability\sep Electromagnetic gyrokinetic simulation\sep particle-in-cell\sep }

\begin{abstract}
\justifying 
The Triangular mesh-based gyrokinetic code TRIMEG-C1 solves the gyrokinetic equations using the particle-in-cell scheme to simulate electromagnetic instabilities in tokamak plasmas. TRIMEG-C1 utilizes a high-order C1 finite element method, which captures the accurate physics with lower grid resolution than the C0 method.
In this work, we focus on achieving a portable implementation on multiple graphics processing unit (GPU) architectures to accelerate the TRIMEG-C1 code for future physics studies. 
The OpenMP framework is chosen as the acceleration framework for GPU offloading on different hardware platforms, specifically, NVIDIA and AMD GPUs. The particle pushing procedure, as well as particle-to-grid operations have been adapted for GPU execution. A speedup of $\approx9$ for the particle pusher kernel is achieved on 2 AMD MI300A APUs (Accelerated Processing Unit) compared with 2 AMD 9754 CPUs. In addition, the efficiency of hybrid MPI-OpenMP offloading parallelization was assessed by oversubscribing GPU resources. The Ion Temperature Gradient (ITG) mode was simulated using the GPU implementation, and its correctness was verified by comparing the physics results in terms of the energy growth rate and the two-dimensional mode structures.
\end{abstract}

\section{\label{sec:introduction}Introduction}
Gyrokinetic particle-in-cell (PIC) simulations have played a crucial role since their early formulation and implementation \cite{lee1983gyrokinetic}. Important physics mechanisms have been identified based on the gyrokinetic PIC simulations in tokamak plasmas, such as the generation of the zonal flows \cite{lin1998turbulent}, the origin of the intrinsic rotation \cite{wang2011trapped}, and the turbulence impact on the edge heat fluxes to the divertors \cite{chang2017fast}. In the identification of new physics,  numerical schemes and physical models have been developed, such as the iterative $p_\|$ scheme \cite{chen2007electromagnetic}, the mixed variable/pullback scheme \cite{mishchenko2019pullback}, and the implicit scheme \cite{lu2021development,sturdevant2021verification}. The Triangular Mesh-based Gyrokinetic code (TRIMEG) has been developed to study micro and macro instabilities with the open field line region taken into account \cite{lu2019development}, with the recent upgrade using the high-order C1 finite element method in unstructured meshes \cite{lu2024gyrokinetic} and the piecewise field-aligned finite element method (PFAFEM)\cite{lu2026PFAFEM}. With the increasing capability of modern supercomputers, multi-scale gyrokinetic simulations covering small electron scales and transport time scales become feasible. 

Recently, GPU offloading has attracted considerable attention due to its significant potential for accelerating large-scale simulations, especially for PIC codes, which expose a high degree of parallelism.
The OpenACC API has been used in the gyrokinetic PIC codes GEM \cite{chen2007electromagnetic}, ORB5 \cite{lanti2020orb5}, and EUTERPE \cite{kleiber2024euterpe}, as well as in the MHD-kinetic code JOREK \cite{hoelzl:hal-04403692}. The gyrokinetic Fortran code GENE employed a more complex approach by implementing kernels with CUDA C++ and an interface with the existing Fortran codebase \cite{germaschewski2021toward}. The choice of OpenACC works well on NVIDIA architectures, which have dominated the accelerator market in the past years; nonetheless, AMD GPUs are increasingly present in newer HPC clusters due to their reduced costs and comparable theoretical performance. However, the support of OpenACC offered by AMD Fortran compilers is almost absent, as the OpenMP framework has been prioritized for GPU offloading. 
The ability to run the code on all available clusters is crucial, since oversubscription of available HPC resources constitutes a significant limitation to running large-scale simulations.
The Max Planck Institute for Plasma Physics (MPIPP) has access, among other systems, to the Viper supercomputer operated by the Max Planck Computing and Data Facility (MPCDF) and based on the AMD MI300A platform. Therefore, portability is a crucial concept in the GPU porting process of the TRIMEG code, even if it entails a compromise on kernel performance. For the above reasons, OpenMP is the only Fortran-compatible approach that enables portability to both NVIDIA and AMD GPU architectures, while minimizing code divergence and structural changes to the existing codebase. Nonetheless, support for OpenMP offered by Fortran compilers is still in the early stages of development, such that advanced programming features, e.g., polymorphism, are not yet supported in the device code. Moreover, some performance optimizations might not be applied, resulting in reduced usage of local data share (shared memory in OpenCL terminology), and runtime behavior might be undefined, requiring extensive debugging to achieve a working solution. 

The paper is organized as follows. 
Section \ref{sec:code_description} briefly describes the features of the TRIMEG-C1 code.
Section \ref{sec:offload} focuses on the GPU programming model and porting of TRIMEG-C1. 
Section \ref{sec:performance-evaluation} reports different performance evaluation analyses. 
Section \ref{sec:results} demonstrates the correctness of the implementation with benchmark simulations.
Section \ref{sec:conclusions} provides a summary and an outlook on future developments.

\section{\label{sec:code_description}Description of the TRIMEG-C1 code}
TRIMEG-C1 has been developed based on the earlier work of TRIMEG-C0, both of which utilize the finite element method (FEM) to solve the field equations on an unstructured triangular mesh. TRIMEG-C0 was originally developed for the simulation of the ion temperature gradient (ITG) mode in the whole tokamak plasma volume using the adiabatic electron model \cite{lu2019development}. It has been recently upgraded with the electromagnetic gyrokinetic model, including kinetic electrons. The C1 version mostly differs in the numerical methods, as it features a scheme based on $C^1$ continuous finite element basis functions \cite{lu2024gyrokinetic}. These are used to represent the discrete variables in the poloidal cross-section, while a cubic B-spline is used in the toroidal direction.
The field equations are solved with \texttt{PETSc}, a library designed for scientific applications modeled by partial differential equations, thus requiring linear algebra tools to solve the discretized system. 
In pushing the particles, the perturbed field is interpolated at each particle position using the field variables on the grid. The resolution of the field equations requires the charge and current densities on the grid, which are calculated by projecting the quantities stored at the particle locations in the continuous phase space onto the finite element grid.

%\subsection{Gyrocenters' equations of motion}
\subsection{Physics equations}
TRIMEG implements the gyrokinetic model, which is valid under the assumption that the timescale of the particle gyromotion around its magnetic axis is much smaller than that of its parallel motion and of field fluctuations. Moreover, the Larmor radius must be small when compared to macroscopic length scales.
This implies a reduction of the phase space from six dimensions (three spatial and three velocity components) to five, as the gyromotion is averaged over and the gyrophase is removed as an independent variable. The parallel component of the vector potential is decomposed to the Hamiltonian part and the symplectic part\cite{mishchenko2014pullback,hatzky2019reduction,lu2024gyrokinetic},
\begin{eqnarray}
    \label{mixed variable}
    \delta A_\|=\delta A_\|^{\rm s}+\delta A_\|^{\rm h}\;\;. 
\end{eqnarray}
The particle dynamics is dictated by the gyrocenters' equations of motion, and the normalized version in cylindrical coordinates is found below. 
\begin{eqnarray}
 \dot{\boldsymbol R}_0 
  &=& u_\parallel {\boldsymbol b}^*_0 + \frac{m\mu}{qB^*_\parallel} {\boldsymbol b}\times\nabla B \;\;, \\
  \dot u_{\parallel,0}
  &= &-\mu {\boldsymbol b}^*_0\cdot \nabla B \;\;,
\\
\delta\dot{\boldsymbol R}
 & = &\frac{{\boldsymbol b}}{B^*_\parallel}\times \nabla \langle \delta\Phi -u_\parallel \delta A_\parallel\rangle 
  -\frac{q_s}{m_s}\langle\delta A^{\rm h}_\parallel\rangle {\boldsymbol b}^*\;\;, 
\\
  \delta \dot u_\parallel
  &= & -\frac{q_s}{m_s} \left({\boldsymbol b}^*\cdot\nabla\langle\delta\Phi-u_\parallel\delta A^{\rm{h}}_\parallel\rangle +\partial_t\langle\delta A_\parallel^{\rm{s}}\rangle \right) 
  -\frac{\mu}{B^*_\parallel}{\boldsymbol b}\times\nabla B\cdot\nabla\langle\delta A_\parallel^{\rm{s}}\rangle \;\;,  
\end{eqnarray}
where $\langle\ldots\rangle$ denotes the gyro average, ${\boldsymbol b}^*={\boldsymbol b}_0^*+\nabla\langle\delta A_\parallel^{\rm s}\rangle\times{\boldsymbol b}/B_\parallel^*$, ${\boldsymbol b}^*_0={\boldsymbol b}+(m_s/q_s)u_\parallel\nabla\times{\boldsymbol b}/B_\parallel^*$, ${\boldsymbol b}={\boldsymbol B}/B$ is the unit vector in the direction of the equilibrium magnetic field, and the parallel velocity coordinate is defined as
\begin{eqnarray}
\label{eq:upara_def}
    u_{||}=v_{||} +\frac{q_s}{m_s}\langle\delta A_{||}^h\rangle\;\;.
\end{eqnarray} The effective magnetic field $B^*$ differs from the physical magnetic field in that it incorporates the effects of the particle's parallel velocity. This allows us to get the effective magnetic field $B_{||}^*$ along the field lines, which is defined as $B_\parallel^*=B+(m_s/q_s)u_\parallel{\boldsymbol b}\cdot(\nabla\times{\boldsymbol b})$. 

The gyrocenters' distribution is decomposed into the equilibrium part and the perturbed part, namely, $f=f_0+\delta f$, where $f_0$ is chosen as the Maxwellian distribution, and $\delta f$ is solved along the trajectories,
\begin{eqnarray}
    \frac{d\delta f}{dt} = -\delta\dot{\boldsymbol R} \cdot\nabla f_0 -\delta \dot u_\parallel\frac{\partial f_0}{\partial u_\|} \;\;,
\end{eqnarray}
where $d/dt=\partial/\partial t+\dot{\boldsymbol R} \cdot\nabla + \dot u_\parallel\partial/\partial u_\|$. The perturbed density and the perturbed parallel current are calculated using $\delta f$ and are used to solve the quasi-neutrality equation and the parallel Amp\`ere's law. The ideal Ohm's law is also solved for $\delta A_\|^{\rm s}$. The details of the equations can be found in the previous TRIMEG papers \cite{lu2024gyrokinetic,lu2026PFAFEM}. 

\subsection{Code Implementation}
\begin{figure}
    \centering
    \includegraphics[width=0.5\linewidth]{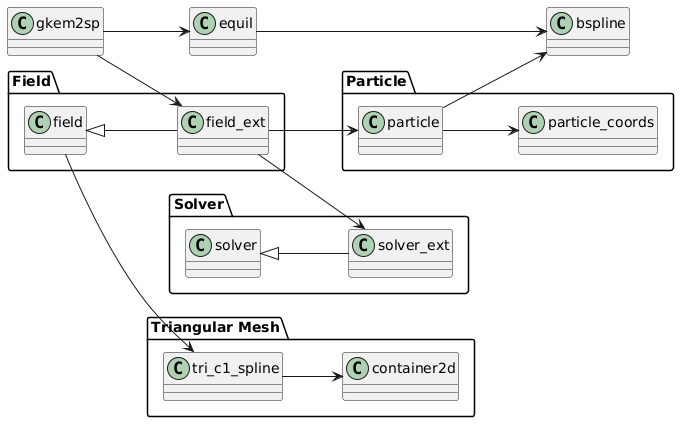}
    \caption{Class Hierarchy of TRIMEG}
    \label{fig:class-uml}
\end{figure}

TRIMEG is implemented in Fortran and makes pervasive use of the object-oriented paradigm (OOP). Figure \ref{fig:class-uml} depicts the hierarchical structure of the classes implemented in the TRIMEG code. The codebase is approximately 15000 lines, excluding external libraries that have been integrated into the code. Specifically, the B-spline interpolation operations rely on an open-source Fortran library, which provides subroutines for 1D--6D interpolation and extrapolation \cite{williams}. Although this avoided the need for a custom in-house implementation, it introduced a dependency on software developed externally, thereby increasing the code complexity and reducing the developer's control. Furthermore, it makes use of modern Fortran programming features, such as polymorphism and type-bound procedures, which are not supported by GPU-compatible compilers. Since the particle pusher kernel contains several calls to B-spline interpolation subroutines, these routines need to be executable on GPU devices. Achieving this was non-trivial and required modifications beyond the library API, which is an undesirable outcome from a software maintainability perspective.

%%%%%%%%%%%%%%%%%%%%%%%%%%%%%%%%%%%%%%%%%%%%%%%%%%%%%%%%%%%%%%%%
\section{GPU Offloading of TRIMEG}
\label{sec:offload}
\subsection{GPU Offloading Paradigm} 
There exist several ways to achieve GPU offloading in Fortran, which vary in their level of abstraction, performance, and portability. Due to the heterogeneity of the available HPC systems, the decision was made to use OpenMP to port TRIMEG to GPU architectures. Specifically, OpenMP supports offloading for both NVIDIA and AMD GPUs, which are part of the Max Planck Computing and Data Facility (MPCDF) Raven and Viper clusters, respectively. The institute also has access to other third-party clusters, such as Pitagora, an NVIDIA-based cluster managed by Cineca.
Similar to OpenACC, OpenMP is a high-level standard that relies mainly on compiler directives, which allow the user to specify the portions of the code to be parallelized and the parallelization strategy to be used. However, although OpenACC support on NVIDIA GPUs is advanced, its support on AMD GPUs is lacking, as this vendor is investing more resources in developing OpenMP offloading support.
Other options for GPU programming in Fortran include CUDA Fortran, Kokkos, and HIP. The first is inherently  not compatible with AMD GPUs, thus requiring the translation of the CUDA kernels to a compatible language, either manually or partially through automated tools, such as GPUFORT, or Large Language Models (LLMs). This approach was avoided to minimize code duplication and because of time constraints.
Kokkos has a C++ core, but a Fortran compatibility layer is available, despite the latter still being in early-stage development. It would require writing Fortran wrappers that call the C++ Kokkos API routines, thus increasing implementation complexity. Moreover, compiler support for Kokkos may be lacking, especially for AMD architectures. The same holds true for HIP.
Given that portability is a key objective of this work, OpenMP was chosen as the GPU programming model. However, it introduces performance limitations due to the programming abstraction. Under the hood, \texttt{omp target} regions are translated into CUDA/HIP kernels, which are not fully optimized. Furthermore, portability across different platforms relies on the compiler's actual level of support for OpenMP offloading, which renders it a potentially fragile solution. Simultaneously developing and testing on both NVIDIA and AMD architectures helps minimize code divergence due to different compiler limitations. 

The main focus of the GPU offloading is on the particle push operation, where the marker particles are pushed along their trajectories in the phase space, and several physical quantities are computed at their locations. This portion of the code is well-suited for GPU offloading as all the particles can be treated and evolved independently, given that there is no interaction between them. This exposes a high degree of parallelism, which can be exploited with the massive number of independent workers on GPU devices. Depending on the input parameter configuration, the particle pushing operation, partially including field-interpolation calculations, can take up to 40\% of the total execution time of a simulation. The remainder is mostly devoted to solving the field equations and projecting particle quantities onto the grid. 

Although OpenMP offloading seems straightforward at first, applying it to complex codes is not a trivial task. While in principle the programmer only needs to inject compiler directives, the actual implementation might require extensive code restructuring and debugging, especially for large kernels. The incomplete support of programming features by Fortran compilers significantly increased the difficulty of GPU porting, and most of the work consisted of circumventing compiler limitations.

\subsection{Particle Push Kernel}
The particle pusher advances marker particles in phase space by integrating the gyrokinetic equations under the influence of electromagnetic fields.
From an implementation perspective, particle pushing is computed in a loop over all marker particles, with each marker evolving independently of the others by solving the gyrocenters' equations of motion. The kernel is characterized by a deep stack of function calls, including interpolation routines of the B-spline library, amounting to a total of $\approx3000$ lines of code that need to be device-callable. In addition, it makes use of a custom implementation of dynamically growing arrays that mimic C++ \texttt{std::vectors}. The functions called from inside the loop body do not contain loops with a large number of iterations, which would otherwise be candidates for an additional layer of parallelism. 

At the beginning of the simulation, a key portion of the particle positioning scheme \cite{lu2019development} is executed, computing the mapping between mesh triangles and rectangular bounding boxes.
The kernel computes the equilibrium configuration of the magnetic field, interpolates the field quantities to the particle positions, solves the gyrocenters' equations of motion, and calculates the weights that determine the contribution of each marker to physical quantities, such as density and current. Five arrays store the updated physical quantities for each marker, namely, the time derivatives of $R$, $Z$, $\varphi$, $v_{||}$, and the marker weights $w$, which need to be transferred from the device back to the host. 

The offloading strategy features  preallocation of array memory on the GPU device, so that the runtime allocation overhead is minimized. Moreover, after the initialization phase, all data structures that remain static during execution are transferred to the GPU once to avoid unnecessary data movements. The entire particle loop is wrapped in an \texttt{!\$omp target teams distribute parallel do} directive, so that it constitutes a single large kernel. This has the advantage of exposing the parallelism across all marker particles, without any loop-carried dependency, thereby allowing a more efficient partitioning of the iteration space into chunks across OpenMP teams, where each chunk is processed in parallel by the GPU threads assigned to each OpenMP team. Moreover, teams and threads need to be spawned only once per pusher call, thus reducing the overhead to a minimum. However, this comes at a substantial implementation cost, since every function inside the loop needs to be callable from the GPU. Furthermore, there is a constraint on the number of marker particles that can be processed on a single node. If too many particles are assigned to a node, its GPU memory will not be sufficient to store all the necessary data at once. Testing simulations showed that processing more than 64 million particles per node leads to Out-Of-Memory (OOM) errors on a node equipped with 2 AMD MI300A APUs, each having 128 GB of High Bandwidth Memory (HBM). 

Register pressure is also a concern. A warp (or wavefront) is the execution unit issued by the hardware scheduler, and is composed by a group of threads (32 or 64) that execute in a lockstep following the Single Instruction, Multiple Thread (SIMT) execution model. A single instruction is fetched for all the threads in the same warp, and each thread executes in on different data. Every thread has its own private context where its state and local variables are stored. If the kernel is large, allocating several local arrays and using many variables, the memory requirements of each thread will sharply increase. The fast on-chip registers are few and might not be sufficient to store the live data of several warps. In such a case, fewer warps can be scheduled, as their memory requirements are high. This results in fewer resident warps per streaming multiprocessor (SM), and a lower hardware occupancy. Therefore, it is important to keep the memory footprint of the kernel as small as possible, and avoid local array allocation where not it is strictly necessary.

\subsection{Other Kernels}
Offloading the particle pusher was the greatest challenge, as it required extensive code refactoring to achieve compatibility with OpenMP offloading. After this porting was completed, the next performance hotspots were the following set of subroutines that perform grid-to-particle (G2P) operations in other sections of the code.
\begin{itemize}
    \item Kernel for pullback treatment. The pullback is a coordinate transformation applied to the perturbed distribution function and is used to map the gyrocenters' mixed variable phase space or $p_\|$ (Hamiltonian) phase space to the $v_\|$ (Symplectic) phase space. If the selected model is electromagnetic and uses either the mixed variable or the Hamiltonian formulation, it is necessary to compute a correction term for the perturbed magnetic potential $\delta A^h$ for each particle marker.
    \item Kernel for density calculation.
    \item Kernel for distribution-function interpolation.
\end{itemize}
Note that all these kernels rely on the same G2P function; therefore, they have nearly identical structures.

\subsection{Code Compilation}
The starting point was to adapt the code so that it could be compiled correctly on both target architectures. During the development phase, it was not trivial to preserve a unified code version due to differences in compiler limitations, highlighting the challenge of achieving portability across both NVIDIA and AMD GPU platforms. For reference, the compiler versions primarily used were 22.0 for \texttt{amdflang}, and 25.1 for \texttt{nvfortran}.

\subsection{Implementation Challenges and Code Restructuring}\label{implementation-problems-code-restructuring}
This section is dedicated to the problems encountered during GPU porting on AMD and NVIDIA architectures. The errors refer to the corresponding compilers, \texttt{amdflang} and \texttt{nvfortran}, respectively. No reference documentation is available that clearly states which features are currently supported by these compilers for OpenMP offloading. Some information may be found on specialized websites, but the overall lack of documentation makes OpenMP offloading with Fortran more challenging than expected. For this reason, compiler limitations are reported in this section, together with workarounds that allow the code to compile and run successfully.

\texttt{amdflang} mostly showed issues during the linking phase. Several linker errors appeared, stating \texttt{undefined reference to function\_name}, referring to Fortran runtime host functions called from within GPU kernels, even the most basic ones, such as integer allocation. The solution is to include the linker flag \texttt{-lflang\_rt.hostdevice}, which instructs the linker to also link device code against Fortran host runtime libraries. In the latest compiler release, version 23.2, this option should no longer be specified explicitly, since it is passed internally to the linker. 
In addition, some functions did not compile correctly, producing the error \texttt{not yet implemented: derived type} for the OpenMP memory-allocation directives. The solution is to use the \texttt{select type} construct on the passed-object dummy argument of the function. This construct is used either to safely downcast from a polymorphic type to a specific derived type or to branch based on the dynamic type of a polymorphic object. At the time of writing, polymorphism is not yet supported by the amdflang compiler, nor are type-bound procedures. The programmer can still choose to use such features, as the code will compile, but the runtime behavior is undefined.

We now discuss the limitations of the \texttt{nvfortran} compiler.
The OpenMP standard requires that a subroutine called from a target region be declared target-callable. However, such a subroutine must also have public scope. This was not the case for the interpolation subroutines of the B-spline library, which exposed only type-bound procedure bindings. Since this feature is not supported, it was necessary to bypass the API and change the scope of the interpolation functions from private to public. 

The error \texttt{error: use of undefined value '@class\_name\_\_\_\_derived\_type\_name\_'} is caused by a stringent limitation of \texttt{nvfortran}, which lacks dynamic typing in this context.
The declaration of the invoking object inside the subroutine, known as the passed-object dummy argument in Fortran jargon, cannot be \texttt{class(derived\_type\_name)} and must instead be changed to \texttt{type(derived\_type\_name)}. This equates to a non-polymorphic approach, since the type of the passed-object dummy argument, i.e., the object used to invoke the function, must match exactly the type  specified in the function definition, and cannot be one that extends it. Moreover, the former declaration is fundamental to enable the use of type-bound procedures, a fundamental feature of extensible data types, in terms of inheritance and polymorphism. Such procedures allow a base implementation to be overridden in an extended derived type, thereby customizing its behavior according to object-specific requirements.
Nevertheless, this feature is rarely used inside TRIMEG, since in most cases it is not necessary to make the derived type extensible. To solve the problem and circumvent this compiler limitation, it is sufficient to remove the extensibility feature from the subroutines to be accelerated. Nonetheless, this imposes a constraint on future developments of the code, so the decision to remove polymorphism should be carefully evaluated. 

Consider the derived types \texttt{bspline\_[1-3]1d}, which extend the abstract type \texttt{bspline} and introduce polymorphic behavior. Objects of these types need to be accessible inside GPU kernels, even though type-bound procedures are not allowed. To ensure correct behavior, the component declaration in the containing derived type, e.g., the equilibrium type, should be changed from \texttt{type(bspline\_[1-3]d)} to \texttt{class(bspline\_[1-3]d), allocatable}, and the component should be allocated accordingly. 

The compiler limitations are not only related to polymorphism. One constraint is that the compiler does not guarantee the consistency of a dynamic array size across function calls. This can generate an illegal memory access error during kernel execution. If the size of an array is declared to be equal to a runtime variable inside a subroutine that is called in the GPU kernel, and that array is passed along a stack of function calls, the programmer must ensure that all its declarations along the chain of function calls specify the actual size, while avoiding the more generic \texttt{dimension(:)}. 

% ============ RUNTIME ISSUES ============
A major issue affected the structure that contains the mapping between bounding boxes and mesh triangles. It is a two-dimensional array of structures (AoS), where the two dimensions correspond to those of the intermediate grid. Each element is a dynamic vector containing all the mesh elements that overlap with the corresponding box. Since standard Fortran does not provide a built-in implementation of dynamically growing arrays, TRIMEG includes a simple library for this purpose. It mimics the behavior of C++ \texttt{std::vector} objects in the standard library and defines a \texttt{VECTOR} object, which is composed of a one-dimensional array called \texttt{data}, where the elements are stored, and a scalar that stores its current size. As previously discussed, the mapping is constructed during the initialization phase of the simulation. The vector structure is needed since the number of triangles overlapping with a single box is unknown a priori. Once the mapping is generated, the vectors do not change, potentially allowing them to be transferred to the GPU only once before time evolution starts. 
The non-trivial mapping of the AoS, which is required on discrete GPUs, is shown below.

\begin{verbatim}
!$omp target enter data map(alloc:tria_vec_obj)
do ftc=1,nx-1
    do fvc=1,ny-1
      !$omp target enter data map(to:tria_vec_obj(ftc,fvc))
      !$omp target enter data map(alloc:tria_vec_obj(ftc,fvc)%data)
      !$omp target update to(tria_vec_obj(ftc,fvc)%data)
    enddo
enddo
\end{verbatim}

where \texttt{nx} and \texttt{ny} are the dimensions of the intermediate grid. 
This set of OpenMP directives is cumbersome and causes runtime issues. In principle, it is sufficient to transfer the data structure only once. However, the OpenMP runtime does not preserve the device pointer to its memory region between different kernel calls, thus generating a CUDA illegal memory access error. The only workaround compatible with this AoS layout consists of allocating the device memory before each kernel call, transferring the data, executing the kernel, and finally deleting the data and deallocating the memory. While this solution works, it is clearly not optimal due to its complexity and the overhead it generates. In the end, the adopted solution was to flatten the AoS into a one-dimensional array after each vector had been populated during the mapping construction, and another one-dimensional array stores the position each vector head in the flattened array. This results in a cleaner mapping that can be performed only once.

\begin{comment}
    
\begin{verbatim}
!$omp target enter data map(alloc:tria_vec_obj_flattened)
!$omp target update to(tria_vec_obj_flattened)
!$omp target enter data map(alloc:tria_vec_obj_index)
!$omp target update to(tria_vec_obj_index)
\end{verbatim}

\end{comment}

After all subroutines and data structures were adapted to run on both architectures, significant numerical errors were present in the output arrays produced by the kernels compared with the CPU results. The cause was a race condition inside the B-spline library, despite the claim that it is thread-safe. The interpolation functions make use of temporary arrays for intermediate calculations. However, these arrays are not declared locally inside the function; rather, they are components of the bspline object. Consequently, they are shared among all threads that call the function in parallel on the GPU. It was sufficient to switch to locally declared arrays, whose size is known only at runtime, and depends on dummy arguments of the \texttt{bspline} object. Fortran allows automatic stack allocation under these conditions, so it was not necessary to declare the arrays as \texttt{allocatable} and explicitly allocate them on the heap.

The GPU implementation exhibited another fault on NVIDIA GPUs, where the simulation would stall during kernel execution, after a seemingly random number of kernel calls. By varying the input parameters, we identified a causal relationship between the GPU grid size, that is, the number of GPU workers processing the workload, and the number of kernel calls that occurred before the stall. Specifically, the larger the grid size, the higher the probability of stalling. This issue required extensive debugging, which eventually traced the root cause to a synchronization bug in the B-spline library. A shared variable was accessed inside a deeply nested subroutine by multiple threads at the same time, eventually causing a deadlock. The code structure had to be changed to avoid this behavior while preserving the semantics, and the problem was solved. It is still unclear why this did not affect the version compiled for the AMD MI300A APU, which was initially misleading. In summary, integrating external code into an existing project implies a lack of full control of its runtime behavior. Moreover, the implementation design of external codes might not be suitable for GPU porting, especially when using a directive-based approach that does not require writing kernels from scratch. The compromise between the simplicity of the GPU programming model and the decreased control over GPU behavior can lead to issues that are hard to debug, and are not reproducible on different architectures due to distinct compiler toolchains. 

\begin{figure}
    \centering
    \includegraphics[width=0.8\linewidth]{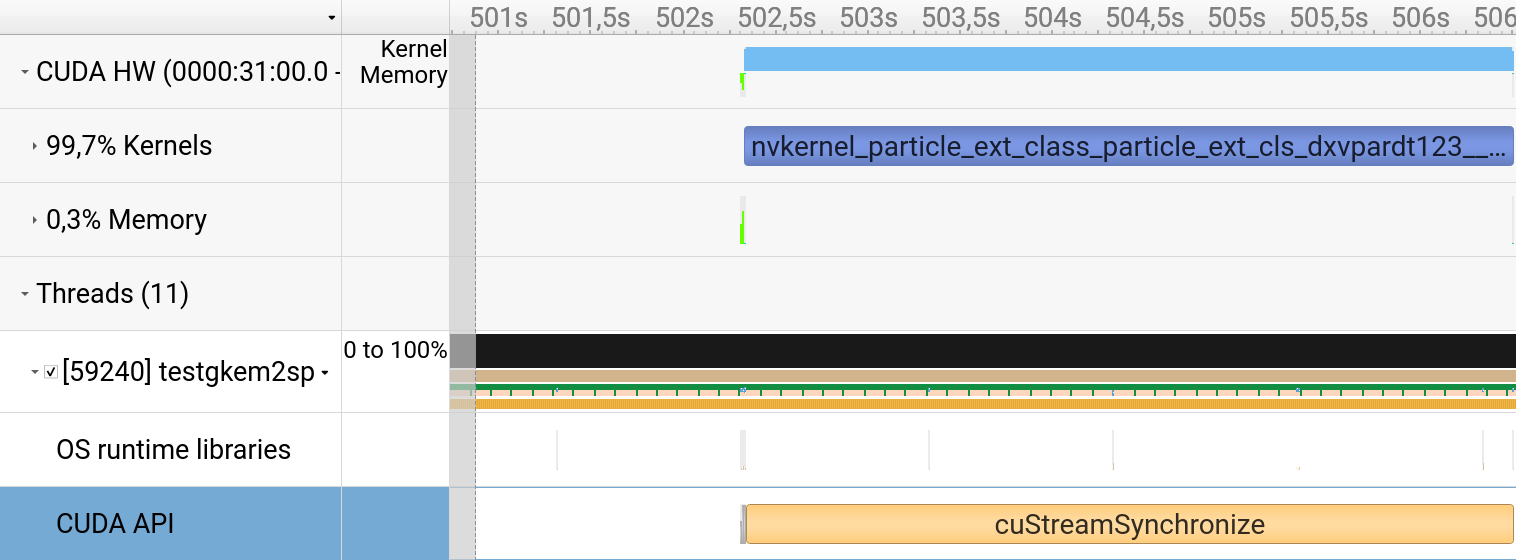}
    \caption{Nsys profiling data of the particle pusher kernel. Memory transfers account only for 0.3\% of the total GPU runtime.}
    \label{fig:nsys-correct-profiling}
\end{figure}

The last major issue was the related to low performance of the kernel on NVIDIA GPUs. The particle pusher kernel was consistently slower than its original CPU version, making GPU porting ineffective. Pinpointing the issue required debugging, manual inlining of functions, and commenting out chunks of code to understand under which conditions the kernel was slowed down. The first possible cause was the cost of explicit memory transfers between the host and the device. There are multiple ways to determine whether this is an actual issue. One is to comment out the entire kernel content except for the memory transfers, and turn off compiler optimizations; otherwise, if a data structure is not accessed inside the kernel, the compiler might avoid transferring it altogether. Another way is to profile the application with \texttt{nsys}, which gathers information about memory movements between host and device, kernel launches, CUDA events, and more. The profiling outcome is shown in Fig. \ref{fig:nsys-correct-profiling}. More than 99.7\% of the GPU execution time is taken up by the kernel itself, which is still extremely slow. Memory transfer events are $\approx3$ orders of magnitude less expensive than kernel execution, which is why they cannot be seen in the figure. The \texttt{cudaStreamSynchronize} event corresponds to the kernel execution and accounts for the synchronization costs of GPU threads when they reach the implicit barrier at the end of the OpenMP target region. This clearly shows that the \texttt{nvfortran} compiler struggles to optimize the kernel, which does not incur such a slowdown when compiled with \texttt{amdflang} with Unified Shared Memory (USM) support. 
Finally, the problem was traced down to the declaration of temporary arrays in the \texttt{evaluate\_[1-3]d} functions of the B-spline library. These arrays had been modified to avoid race conditions, as previously described, and are allocated automatically inside the function. When one of these functions is called inside the GPU kernel, each thread allocates its own local array instance, whose size is known only at runtime and depends on the order of the B-spline functions provided in the input file. This generates substantial overhead, even though the reason why this happens only on NVIDIA platforms is unclear at the time of writing. The solution was to switch to a compile-time size by taking the maximum possible size, which is 12 for the largest array, and does not generate significant performance degradation in terms of register pressure, as the size of the arrays is small. To avoid reducing the flexibility of the code in terms of input parameters, this modification is applied only when compiling for NVIDIA architectures with GPU support, otherwise the original implementation is used. The downside is code divergence, as portability proves once again to depend on the chosen platform, rather than on the GPU programming paradigm. 

\subsection{Kernel Correctness}
After completing the GPU porting of the particle push kernel and the G2P functions therein, it was necessary to assess the correctness of the output results. As previously mentioned, the kernel computes five output arrays that store particle information, namely the time derivatives of the spatial coordinates in cylindrical coordinates, $(R,Z,\varphi)$, of the parallel velocity $v_{||}$, and of the weight coefficient $w$. To obtain an exact comparison between the CPU and GPU results, the kernel is executed on both the host and the device within the same program run, and the outputs are compared element-wise. This allows the kernel to be tested in isolation and independently of the pseudorandomness present in other portions of the code.
As expected, the absolute value of the difference was greater than the machine epsilon, which for a double-precision variable is $\approx 2.22\times10^{-16}$. On AMD platforms, the difference was on the order of $10^{-14}$. Simply disabling Fused Multiply-Add (FMA) instructions with the compiler option \texttt{-ffp-contract=off} demonstrated that the kernel is semantically correct, reducing the differences to machine precision. Note that double-precision floating-point values are necessary for accurate and high-fidelity gyrokinetic plasma simulations.

However, the corresponding option for \texttt{nvfortran}, which is \texttt{-Mnofma}, did not remove all discrepancies on NVIDIA platforms.
The debugging tool \texttt{racecheck}, part of the \texttt{compute-sanitizer} suite, was used to automatically search for race conditions, but none were detected. Manual inspection yielded the same result; all functions causing numerical differences were identified and found not to contain any race conditions. The conclusion is that some arithmetic operations have different hardware implementations on GPU architectures, and the OpenMP abstraction makes it extremely difficult to expose the low-level transformations that the compiler applies to device code. The GPU version of TRIMEG is still capable of reproducing physically relevant results, but agreement with the CPU version can be hindered by several factors, such as low plasma pressure in simulations, which are less numerically stable. The actual agreement of physical results will be discussed in Section \ref{sec:results}.

%%%%%%%%%%%%%%%%%%%%%%%%%%%%%%%%%%%%%%%%%%%%%%%%%%%%%%%%%%%%%%%%
%\subsection{Implementation Challenges}
%\label{subsec:racing}

\section{Performance Evaluation}
\label{sec:performance-evaluation}%
\graphicspath{Images/}

Evaluating the performance of the kernels is of paramount importance, given that the purpose of GPU porting is to accelerate the code. There are several ways to compare the speed of the CPU and GPU versions, and most of them depend on a set of assumptions, as there is no universal strategy for doing so. 
First, we analyzed each kernel individually, using an increasing number of MPI processes on a single node. A hybrid MPI-OpenMP offloading strategy is fundamental for running simulations with TRIMEG, which scales very well with the number of MPI processes. Note that the code does not use OpenMP multithreading, so each CPU core should be mapped to an MPI task to fully utilize the hardware resources. We evaluated the performance degradation due to oversubscription of GPU resources as a function of the number of MPI processes, to determine whether hybrid parallelization can work efficiently for large-scale simulations, as the CPU-bound portion requires a high degree of CPU parallelism.

For reference, the hardware capabilities of the clusters used for testing and production runs are listed in Tables~\ref{tab:gpu_clusters} and \ref{tab:cpu_cluster}. 
TOK (node equipped with two AMD EPYC 9754) is a powerful CPU-only HPC cluster, optimized for high-throughput parallel scientific workloads. Pitagora (dual Xeon 6548Y + 4× H100) is a top-tier GPU-accelerated node delivering strong performance for AI and hybrid HPC simulations. Raven (dual Xeon 8360Y + 4× A100) is a strong and cost-efficient GPU HPC node for scientific and AI workloads. Viper (dual AMD MI300A APU) is an advanced unified CPU–GPU system with high memory bandwidth, ideal for tightly coupled scientific computing.

\begin{comment}
The Viper GPU cluster is equipped with two AMD MI300A APUs per node, each with 24 cores, 128 GB of HBM3, 228 compute units, 14592 stream processors, and a theoretical performance of 61.3 TFLOPS for FP64 vector operations. \\
The NVIDIA-based clusters, Raven and Pitagora, feature different GPU architectures. A Raven GPU node is equipped with four NVIDIA A100 GPUs, each with 40 GB of HBM2. Both the CPU and GPU nodes are equipped with an Intel Xeon IceLake Platinum 8360Y with 36 cores. \\
A Pitagora booster node consists of four H100 SXM 80GB HBM2e, and 2 Intel Emerald Rapids 6548Y 2.4 GHz CPUs with 32 cores each, for a total of 64.
The CPU-based cluster is TOK, where a single node consists of two AMD EPYC 9754, with 128 cores/CPU, and a maximum boost clock of 3.1 GHz. \\
\end{comment}

% ================= GPU Clusters =================
\begin{table}[h]
\centering
\small
\begin{tabular}{|l|l|l|l|}
\hline
\textbf{Cluster} & \textbf{CPU} & \textbf{GPU} & \textbf{Key Specs} \\ \hline

Viper 
& -- 
& 2× AMD MI300A APU (128 GB)
& 24 cores/APU \\ \hline

Raven 
& Intel Xeon 8360Y
& 4× NVIDIA A100 (40 GB) 
& 36 cores \\ \hline

Pitagora 
& 2× Intel 6548Y
& 4× NVIDIA H100 (94 GB) 
& 72 cores/CPU \\ \hline

\end{tabular}
\caption{GPU-accelerated clusters.}
\label{tab:gpu_clusters}
\end{table}

% ================= CPU Cluster =================
\begin{table}[h]
\centering
\small
\begin{tabular}{|l|l|l|}
\hline
\textbf{Cluster} & \textbf{CPU} & \textbf{Key Specs} \\ \hline

TOK 
& 2× AMD EPYC 9754
& 128 cores/CPU \\ \hline

\end{tabular}
\caption{CPU-based cluster.}
\label{tab:cpu_cluster}
\end{table}

The concept of an APU is different from that of a discrete GPU in that it features a physically unified shared memory space between the host and the device. USM can be harnessed, thus avoiding redundant data copies and relying on page faults to map the data into the device's virtual memory. A page fault is an interrupt that occurs when a kernel tries to access a virtual memory page that is not currently mapped in the GPU page tables, a data structure translating virtual addresses into physical ones. For an APU, a page fault does not trigger a page migration, as the data is already present in the physical memory, but only a page table population. The USM introduces an additional layer of abstraction that does not require explicit data transfer directives, similar to using \texttt{-gpu=mem:managed} on NVIDIA GPUs, and can improve kernel performance. Note that pointers to allocatable memory still require manual mapping.

In the performance analyses, we compared individual kernel execution times on the CPU and GPU, as well as overall code performance. The CPU baseline was obtained by running the code on the CPU TOK cluster, even though this approach entails comparing kernel performance on fundamentally different hardware configurations. Ultimately, it is not possible to achieve a fully fair comparison, which is why the speedup figures must be interpreted with caution. We used only optimizations that do not involve aggressive arithmetic approximations, which would compromise the necessary level of precision.

\subsection{Individual Kernel Analysis} \label{individual-kernel-analysis}
First, we evaluated the performance of each kernel in isolation. This is useful for understanding whether the workload is suitable for GPU acceleration, and for quantifying the speedup achievable under ideal conditions, i.e., each MPI process targets a single GPU and can fully utilize its resources. However, in realistic scenarios, TRIMEG needs to allocate as many MPI processes as cores on each node, as the available clusters are oversubscribed and allocating several nodes is often not a viable option. For this reason, an assessment of performance degradation was carried out when oversubscribing each GPU device, that is, using an increasing number of MPI processes on a single GPU node. Each MPI process is associated with a separate context on the GPU, and the GPU driver serializes their execution streams through time slicing. Kernels are interleaved, and context switches can generate non-negligible overhead, thus reducing the overall performance. Moreover, memory transfers are fragmented and serialized on the PCI Express bus that connects the host to the device, lowering the effective bandwidth. All these factors need to be considered when opting for GPU acceleration of a code that is parallelized only with MPI, where CPU cores cannot be mapped to OpenMP threads. 

Table \ref{pusher-kernel-hybrid-parallelism-tok} shows the speedup of the particle pusher kernel on the Viper GPU node with USM enabled with respect to the original CPU implementation on the TOK cluster. The workload size is $10^6$ marker particles, which is small for a full node, but is computationally expensive to process on a single core. A reduced performance study using a larger workload is presented subsequently. The kernel execution times are obtained as the average over a number of runs equal to $\rm{\#MPI\_procs}\cdot 20$, for both GPU and CPU runs. The first and second rows correspond to ideal GPU execution conditions, where each MPI process can fully use the corresponding GPU. Even though the size of the kernel is halved when the number of MPI processes is doubled on the same GPU node, as the workload is distributed across MPI ranks, a reduction in execution time is not expected, as kernels are serialized on the accelerator. The only advantage is obtained by using at least two MPI processes, so that both APUs on the node are utilized, and consequently the average kernel execution time is approximately halved. Non-Uniform Memory Access (NUMA) effects are avoided by using the \texttt{--gpu-bind=closest} Slurm option, which automatically binds the MPI ranks on a socket to the APU on that same socket. The CPU version exhibits excellent scaling when increasing the number of MPI processes, progressively reducing the advantage of using hybrid MPI/OpenMP offloading acceleration. It should be noted that MPI parallelism is more than tenfold when using a TOK with 256 cores compared to 24 cores on Viper GPU, even though single-core performance is lower. Moreover, it is not possible to allocate more than 24 MPI processes on a single GPU node, even though up to 48 are available. Each MPI process creates its own context on the GPU, requiring a significant amount of memory, eventually causing OOM errors when oversubscription is too extreme. Nonetheless, the hardware configuration of Viper allows oversubscription of GPU resources without incurring performance degradation, even when the number of MPI tasks is increased to the maximum.

The same comparison was carried out for the kernels that contribute to the pullback treatment, particle density calculation, and distribution function interpolation, and the results are shown in Table \ref{calcN-pullback-kernel-hybrid-parallelism-tok}. They have been aggregated because the structure of the kernels is the same.
The speedup figures are lower because the number of operations is smaller, reducing the ratio of arithmetic operations to bytes mapped to the device addressing space. Again, execution on a CPU-based cluster drastically reduces the speedup of the GPU-accelerated implementation when allocating all the available cores. 

Then, we conducted a reduced analysis on the Raven and Pitagora NVIDIA-accelerated clusters.
Table \ref{pusher-kernel-hybrid-parallelism-raven} shows the results of the particle pusher kernel on a Raven GPU node compared with a TOK CPU node. The first row corresponds to the performance achieved with one A100 GPU on a GPU node, and one MPI process for both CPU and GPU versions. The fastest kernel time is achieved with 4 MPI processes, one per GPU, with a speedup of $\approx 3.2$ compared to using only one. The last row corresponds to testing the implementations on Raven with 36 MPI processes, whereas the maximum of 256 MPI processes is used on TOK. Again, it was not possible to spawn 72 MPI processes due to OOM errors. Increasing the number of MPI processes from 4 to 36 moderately reduces the performance, causing the kernel to be $\approx 35\%$ slower. Moreover, the average execution time on this hardware configuration is slower than that on the Viper GPU node by $\approx 83\%$ on a full node. However, comparing the fastest GPU configuration, that is, 4 cores per node, against TOK with 256 cores yields a speedup of $\approx5.2$. The trade-off is that the overall code would have only $1/9$th of the MPI parallelism, causing the CPU-bound parts to be extremely slower, thus nullifying the advantage brought by the GPU acceleration of the particle pusher.

Analogously, the performance of the pusher kernel on a Pitagora GPU node compared with a TOK CPU node is reported in Table \ref{pusher-kernel-hybrid-parallelism-pitagora}. The same methodology is used for the performance comparison. When allocating a full GPU node on Pitagora, 64 MPI processes and four H100 GPUs are used. The computational power of this chip is significantly higher than that of its predecessor, the A100. When allocating all available cores, the kernel achieves a speedup of $\approx2.8$ on a Pitagora node compared to a Raven node, and a speedup of $\approx1.5$ compared to a Viper GPU. The fastest setup is obtained when spawning 32 MPI processes, which corresponds to a potential speedup of $\approx14$ when compared to a full TOK CPU node, the trade-off being half of the MPI parallelism.\\

As before, the speedup of the three smaller kernels is aggregated, and the performance of the GPU implementation on both Raven and Pitagora is compared against the CPU version on TOK. The results are shown in Tables \ref{other-kernels-hybrid-parallelism-raven} and \ref{other-kernels-hybrid-parallelism-pitagora}. The advantage of GPU acceleration is almost absent when compared to a full CPU node, due to the limited workload size and because data transfers account for a large portion of the total execution time.

We now consider a larger simulation with $32\cdot10^{6}$ electrons on a full node, which is on the order of magnitude required to achieve realistic particle-in-cell simulations with TRIMEG.
We analyzed the performance on both Viper GPU and Pitagora, using TOK as the CPU baseline, and the results are reported in Table \ref{pusher-kernel-hybrid-parallelism-big-case}. The speedup achieved by the pusher kernel is higher than in the previous analysis due to increased resource occupancy on the GPU nodes.

The same analysis was carried out for the three smaller kernels. As shown in Table \ref{other-kernels-hybrid-parallelism-big-case}, a Pitagora GPU node outperforms a Viper GPU node by $\approx 60\%$. Increasing the workload size proved especially beneficial for the smaller kernels, resulting in a significant speedup.

\begin{table}[H]
\centering
\begin{tabular}{|cccc|}
\hline 
\rowcolor{bluePoli!40}
\#MPI Procs & Viper GPU (s) & TOK CPU (s) & Speedup \\
\hline \hline
1  & 0.0478 & 29.65 & 620 \\
2  & 0.0265 & 15.99 & 603 \\
4  & 0.0256 & 7.79 & 304 \\
8  & 0.0269 & 3.71 & 138 \\
16 & 0.0259 & 1.89 & 73 \\
max* & 0.0215 & 0.123 & 5.7 \\
\hline
\end{tabular}
\\[3pt]
*24 tasks on a GPU node, 256 on a CPU node
\caption{Pusher kernel speedup on AMD-based Viper GPU with USM with respect to CPU version on TOK, problem size of $10^6$ electrons.}
\label{pusher-kernel-hybrid-parallelism-tok}
\end{table}

\vspace{-15pt}

\begin{table}[H]
\centering
\begin{tabular}{|cccc|}
\hline
\rowcolor{bluePoli!40}
\#MPI Procs & Viper GPU (s) & TOK CPU (s) & Speedup \\
\hline \hline
1  & 0.00677 & 4.02 & 594 \\
8  & 0.00395 &  0.509 & 129 \\
16  & 0.00413 &  0.255 & 62 \\
max* & 0.00346 & 0.015 & 4.3 \\
\hline
\end{tabular}
\\[3pt]
*24 tasks on a GPU node, 256 on a CPU node
\caption{Speedup of the kernels for pullback, density calculation and distribution function interpolation on AMD-based Viper GPU with USM with respect to CPU version on TOK, problem size of $10^6$ electrons.}
\label{calcN-pullback-kernel-hybrid-parallelism-tok}
\end{table}

\vspace{-15pt}

% ============ RAVEN DATA ============ %
\begin{table}[H]
\centering
\begin{tabular}{|cccc|}
\hline
\rowcolor{bluePoli!40}
\#MPI Procs & Raven GPU (s) & TOK CPU (s) & Speedup \\
\hline \hline
1 & 0.0743 & 29.65 & 399 \\
4 & 0.0235 & 7.79 & 331 \\
16 & 0.0292 & 1.89 & 65 \\
max* & 0.0393 & 0.123 & 3.1 \\
\hline
\end{tabular}
\\[3pt]
*36 tasks on a GPU node, 256 on a CPU node
\caption{Pusher kernel speedup on a Raven NVIDIA GPU node with respect to CPU version on TOK, problem size is $10^6$ electrons.}
\label{pusher-kernel-hybrid-parallelism-raven}
\end{table}

\vspace{-15pt}

\begin{table}[H]
\centering
\begin{tabular}{|cccc|}
\hline
\rowcolor{bluePoli!40}
\#MPI Procs & Raven GPU (s) & TOK CPU (s) & Speedup \\
\hline \hline
1  & 0.0266 & 4.02 & 151 \\
4  & 0.0112 & 1.07 & 96 \\
% 8  & 0.0174 &  0.509 & 29 \\
16  & 0.0136 &  0.255 & 18.8 \\
max* & 0.0120 & 0.015 & 1.25 \\
\hline 
\end{tabular}
\\[3pt]
*36 tasks on a GPU node, 256 on a CPU node
\caption{Speedup of the kernels for pullback, density calculation, and distribution function interpolation on a Raven NVIDIA GPU node with respect to CPU version on TOK, problem size is $10^6$ electrons.}
\label{other-kernels-hybrid-parallelism-raven}
\end{table}

\vspace{-15pt}

% ============ PITAGORA DATA ============ %
\begin{table}[H]
\centering
\begin{tabular}{|cccc|}
\rowcolor{bluePoli!40}
\hline
\#MPI Procs & Pitagora GPU (s) & TOK CPU (s) & Speedup \\
\hline \hline
1  & 0.0399 & 29.65 & 743 \\
16  & 0.0102 & 1.89 & 185 \\
32  & 0.0088 & 0.955 & 109 \\
max* & 0.014 & 0.123 & 8.8 \\
\hline
\end{tabular}
\\[3pt]
*64 tasks on a GPU node, 256 on a CPU node
\caption{Pusher kernel speedup on a Pitagora NVIDIA GPU node with respect to CPU version on TOK, problem size is $10^6$ electrons.}
\label{pusher-kernel-hybrid-parallelism-pitagora}
\end{table}

\vspace{-15pt}

\begin{table}[H]
\centering
\begin{tabular}{|cccc|}
\hline
\rowcolor{bluePoli!40}
\#MPI Procs & Pitagora GPU (s) & TOK CPU (s) & Speedup \\
\hline \hline
1  & 0.00968 & 4.02 & 415 \\
16  & 0.00511 & 0.226 & 44.2 \\
32  & 0.0074 & 0.114 & 15.4 \\
max* &  0.00841 & 0.015 & 1.8 \\
\hline 
\end{tabular}
\\[3pt]
*64 tasks on a GPU node, 256 on a CPU node 
\caption{Speedup of the kernels for pullback, density calculation, and distribution function interpolation on a Pitagora NVIDIA GPU node with respect to CPU version on TOK, problem size is $10^6$ electrons.}
\label{other-kernels-hybrid-parallelism-pitagora}
\end{table}

\vspace{-15pt}

% ============ BIG SIM DATA ============ %
\begin{table}[H]
\centering
\begin{tabular}{|ccccc|}
\hline
\rowcolor{bluePoli!40}
TOK CPU (s) & Viper GPU (s) & Speedup & Pitagora GPU (s) & Speedup \\
\hline \hline
3.22 & 0.363 & 8.9 & 0.357 & 9 \\
\hline
\end{tabular}
\caption{Pusher kernel speedup on AMD-based Viper GPU and NVIDIA-based Pitagora with respect to CPU version on TOK. The problem size has been increased to $32\cdot10^{6}$ electrons.}
\label{pusher-kernel-hybrid-parallelism-big-case}
\end{table}

\vspace{-15pt}

\begin{table}[H]
\centering
\begin{tabular}{|ccccc|}
\hline
\rowcolor{bluePoli!40}
TOK CPU (s) & Viper GPU (s) & Speedup & Pitagora GPU (s) & Speedup \\
\hline \hline
0.377 & 0.044 & 8.6 & 0.0272 & 13.8 \\
\hline
\end{tabular}
\caption{Speedup of the kernels for pullback, density calculation, and distribution function interpolation on AMD-based Viper GPU and NVIDIA-based Pitagora with respect to the CPU version on TOK. The problem size has been increased to $32\cdot10^{6}$ electrons.}
\label{other-kernels-hybrid-parallelism-big-case}
\end{table}

%\newpage
\subsection{Strong Scalability}
\begin{figure}[H]
    \centering
    \includegraphics[width=0.6\linewidth]{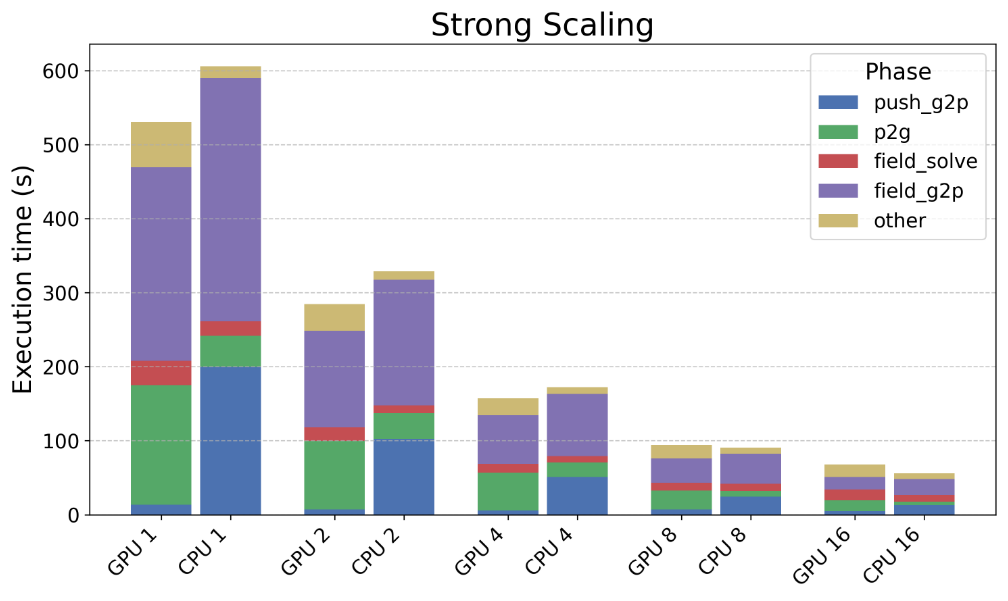}
    \caption{Strong scaling analysis conducted on 1, 2, 4, 8, and 16 GPU nodes compared against CPU nodes. The GPU implementation was run on AMD-based Viper GPU, while the CPU version was run on TOK. The pusher kernel corresponds to \texttt{push\_g2p} and shows significant speedup and good scaling when increasing GPU nodes. }
    \label{fig:strong-scaling-tok}
\end{figure}

Strong scalability aims to evaluate how much a fixed workload can be accelerated when more processing power is added. This is primarily a benchmark for MPI parallelization and might seem out of scope for evaluating the GPU acceleration. However, this analysis should confirm that GPU offloading can work efficiently together with MPI on multiple nodes and may provide insights into new code bottlenecks. We evaluated the strong scalability of a simulation with $32\cdot 10^6$ electrons and $10^6$ ions for an increasing number of nodes, up to 16. Prior to running the simulation, it is important to verify whether register spilling occurs for such a large workload, and, if necessary, reduce the block size to decrease scalar register and vector general purpose register (VGPR) usage.
A strong scaling analysis of the GPU implementation compared with the CPU version executed on TOK is shown in Fig. \ref{fig:strong-scaling-tok}.
The simulation was run for only 10 time steps and was set up to be economical, which is why the particle pushing procedure is not dominant. Achieving meaningful physical results requires a number of time steps on the order of thousands, and different input parameter configurations result in the pusher being more expensive overall, so that the execution times of the different phases would shift further in favor of the GPU-accelerated implementation.

The performance gain of the accelerated code is marginal, primarily because the \texttt{p2g} phase strongly benefits from the 256 MPI tasks available on each CPU node. For this reason, the advantage decreases for an increasing number of nodes, and the overall performance of the GPU implementation is comparable to that of the CPU version, even though there are notable differences in time spent in the different phases. Nevertheless, the result confirms the performance gain of the particle pushing procedure due to GPU acceleration, and the kernel performance does not significantly degrade as the number of nodes increases.

%\clearpage
\subsection{Profiling Data} \label{profiling-data}
\begin{figure}[H]
    \centering
    \includegraphics[width=\linewidth]{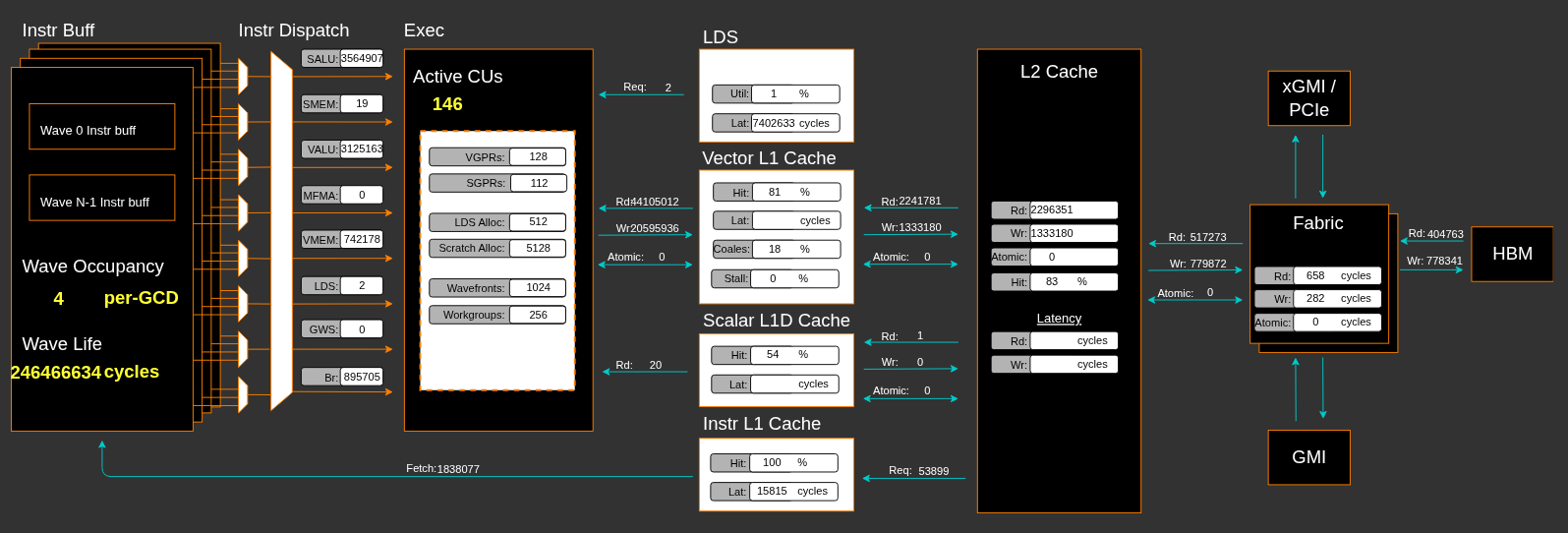}
    \caption{Overview of the GPU architecture, hardware utilization, including CUs, wavefronts and workgroups, dispatched instruction types, and amount of data transfers between the different memory levels for the pusher GPU implementation.}
    \label{fig:overall-dxvpardt-kernel-profiling}
\end{figure}

\begin{figure}[H]
    \centering
    \includegraphics[width=\linewidth]{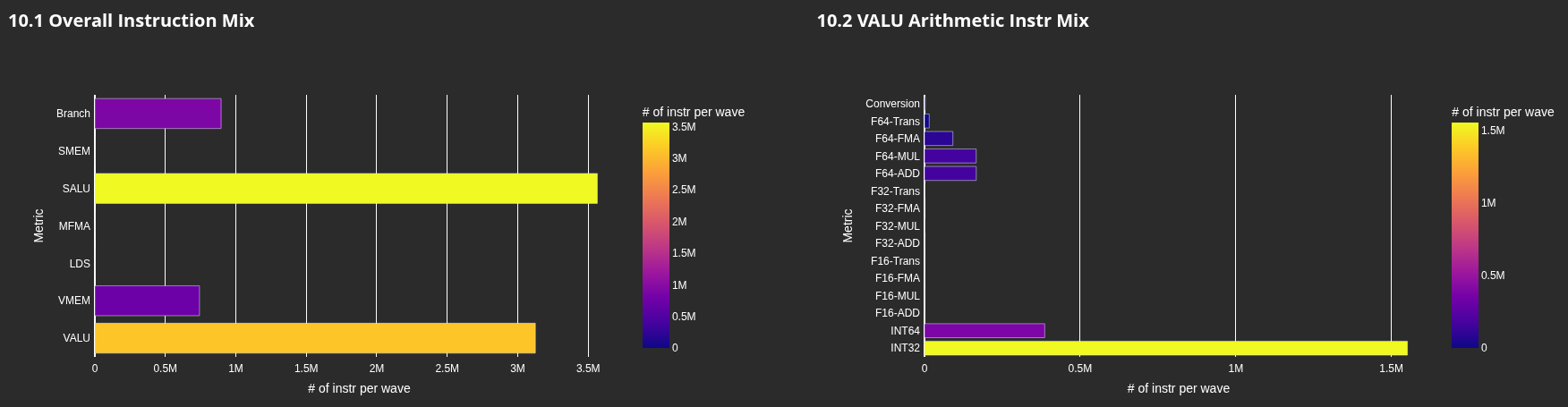}
    \caption{Overall and VALU instruction mix per wavefront for the pusher GPU implementation.}
    \label{fig:overll-instruction-mix}
\end{figure}

Detailed information on kernel performance can be generated with profiling tools specifically provided by hardware vendors. Several metrics can be collected, such as bytes read and written to the global memory, L1 and L2 cache hit rates, the number of vector and scalar instructions executed, and hardware-resource occupancy. \texttt{rocprof-compute} is the tool developed to profile kernel behavior on AMD GPU devices. We profiled a simulation of $4\cdot10^6$ electrons and $10^6$ ions evolved for a single time step, which is sufficient to run all kernels at least once. It is important to keep the computational cost of the simulation low, since collecting all the performance and utilization metrics requires running the program several times. For the sake of brevity and to avoid redundancy, we report only profiling data for the particle push kernel, which is the most computationally expensive one. Fig. \ref{fig:overall-dxvpardt-kernel-profiling} gives an overview of GPU resource utilization and data transfers for the particle pusher kernel. The memory hierarchy is clearly represented, as well as the flow of data. The number of active CUs is only 146 out of the 228 available on the MI300A, due to high register pressure, as each thread creates a private copy of all the local arrays allocated inside the kernel. Note that the Local Data Share (LDS), analogous to shared memory on NVIDIA GPUs, is almost not used. However, not much can be done in this respect, as the OpenMP API does not allow explicit use of it. This represents a potentially important limitation, since LDS is a low-latency, high-performance storage resource.
Fig. \ref{fig:overll-instruction-mix} shows the overall and Vector Arithmetic Logic Unit (VALU) instruction mix for the particle push kernel. Note that all threads in a wavefront execute the same instruction at each clock cycle in a SIMD manner, which is a fundamental feature for achieving good performance and power efficiency. The dominant instruction types are Scalar ALU (SALU) and VALU instructions, which amount to $\approx80\%$ of the total, whereas branches and Virtual Memory Instructions (VMEM) are less common. This is important for achieving high arithmetic throughput and is consistent with high arithmetic intensity, a fundamental metric for evaluating the computational requirements of an algorithm. VALU instructions are primarily issued for integer operations, while FP64 instructions, corresponding to the Fortran type \texttt{real*8}, are much fewer.

Memory operations incur much higher latency than arithmetic instructions; therefore, optimizing them is of paramount importance. The L1 vector data cache hit rate averages 80.51\%, and the L2 hit rate averages 83.12\%. Assuming that the L2 hit rate is measured for L1 misses, this implies that only $(1-0.8051)\cdot(1-0.8312)\approx0.033$, or $3.3\%$ of the read and write operations access global memory, indicating an effective utilization of the cache hierarchy. Only 18\% of the accesses are coalesced, but the memory access patterns in the kernel cannot be easily changed to increase this metric. A structure of array (SoA) is used to store particle data, and is the most suitable for GPU memory access patterns. The scalar data cache is almost unused, as shown in Fig. \ref{fig:overall-dxvpardt-kernel-profiling}. Overall, the memory access patterns appear to be quite efficient because of high locality and data reuse.

\newpage
\section{Numerical Results}
\label{sec:results}
%\label{chap:numerical_results}
The Cyclone base case \cite{gorler2016intercode} and the TCV case are considered in this section to verify the GPU offloading. For both cases, the equilibrium density and temperature profiles, denoted generically by $H(\tilde r)$, are prescribed through 
\begin{eqnarray}
\label{eq:nTprofile}
     \frac{H(\tilde r)}{H(\tilde r_0)}  =\exp\left[-\kappa_\mathrm{H}w_\mathrm{H}\frac{a}{R_0}\tanh\left(\frac{\tilde r-\tilde r_0}{w_\mathrm{H}a}\right)\right]\;\;,  
\end{eqnarray}
where the radial coordinate is defined as $\tilde r=\sqrt{(R-R_{\rm axis})^2+(Z-Z_{\rm axis})^2}$, $L_\mathrm{H}=-\left[{\rm d}\ln H(\tilde r)/{\rm d}\tilde r\right]^{-1}$ is the characteristic length of the profile $H(\tilde r)$, and $\tilde r_0=a/2$. The remaining parameters are introduced in Sections~\ref{subsec:CBC} and \ref{subsec:tcv}. 

\subsection{Study of the Cyclone Case}
\label{subsec:CBC}
\begin{figure}[t]
    \centering

    \includegraphics[width=0.6\linewidth]{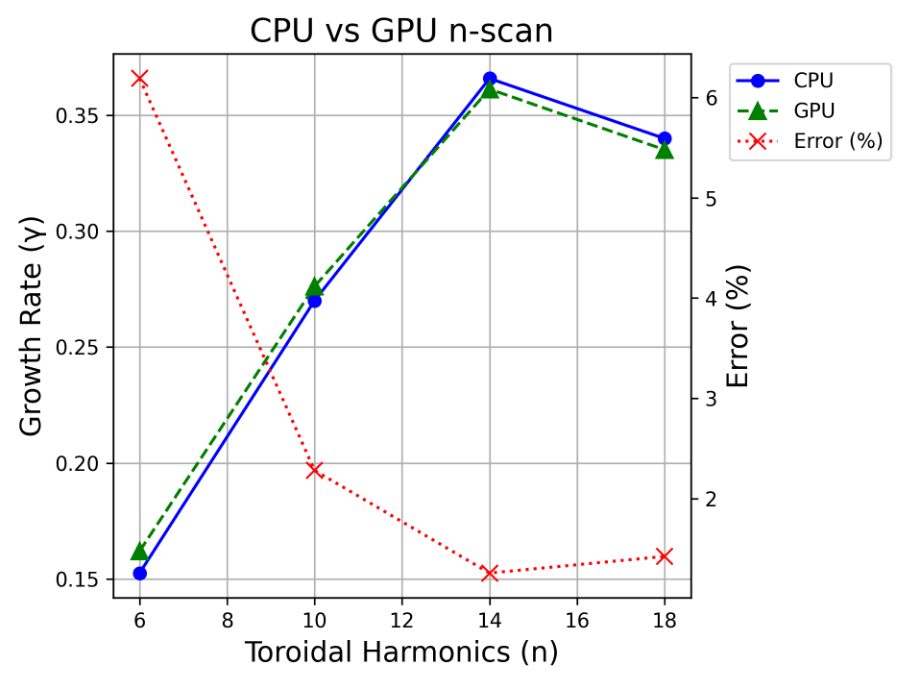}
    \caption{Energy growth rate comparison between CPU and GPU versions for toroidal mode numbers $n=6, 10, 14, 18$. The error is within 6\% for $n=6$, and consistently under 3\% for higher toroidal mode numbers.}
    \label{fig:growth-rate-comparison}

    \vspace{0.5em}

    \includegraphics[width=0.48\linewidth]{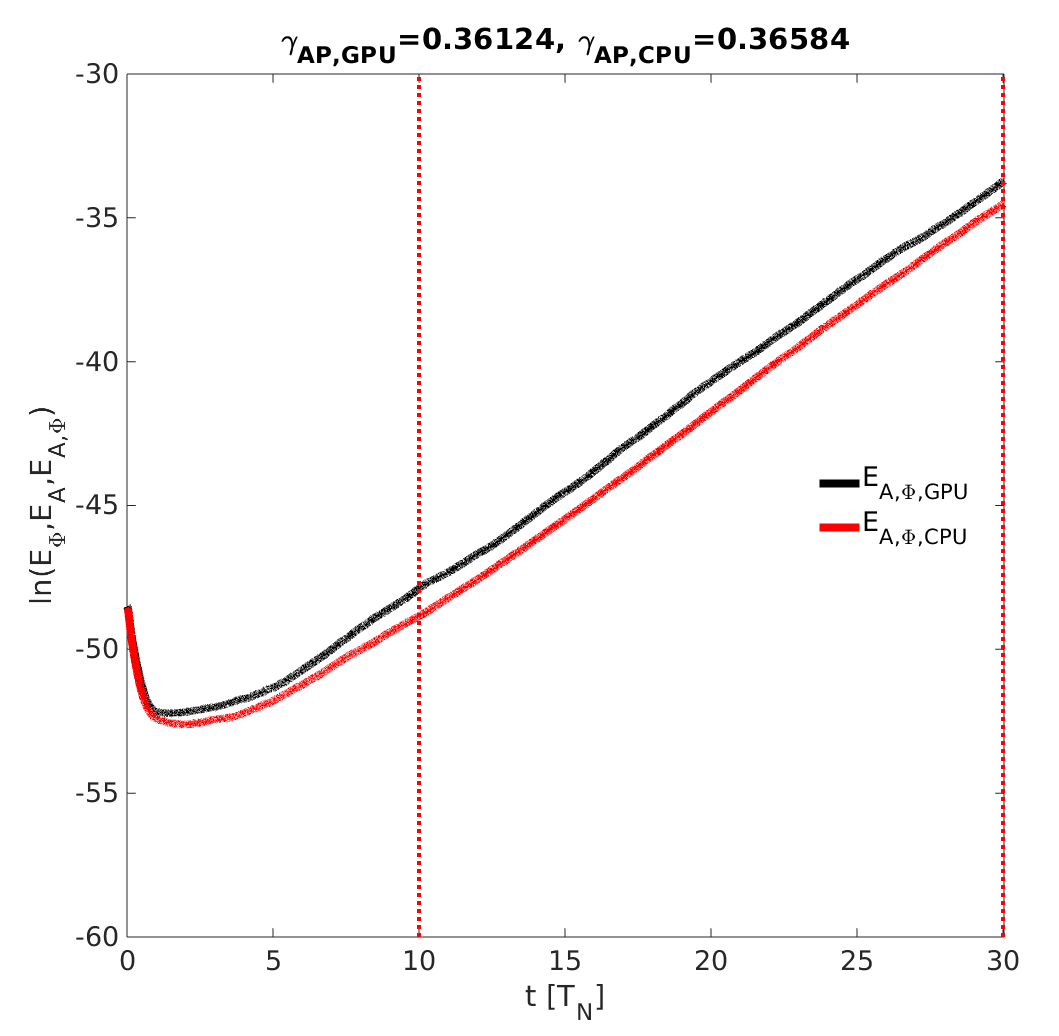}
    \caption{Comparison of the CPU and GPU time evolution of the total energy stored in the electromagnetic field for $n=14$. The energy growth rate corresponds to the selected time interval [10, 30] of the linear stage.}
    \label{fig:total-energy-evolution}

    \vspace{0.5em}

    \includegraphics[width=.48\linewidth]{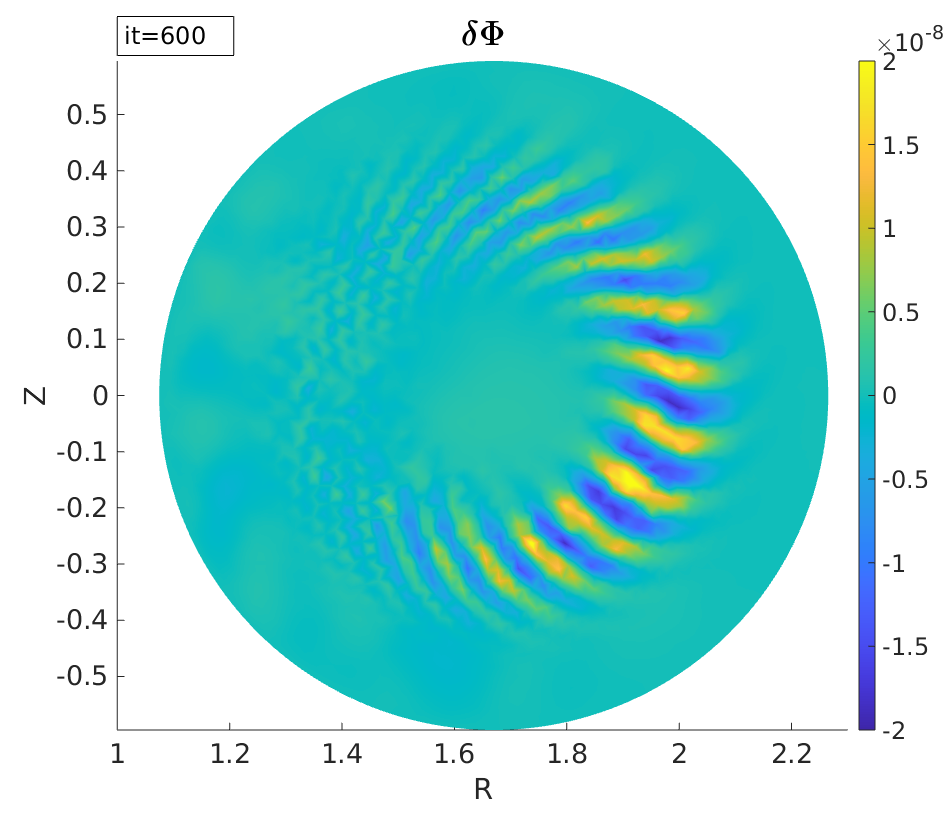}
    \hspace{0.02\linewidth}
    \includegraphics[width=.48\linewidth]{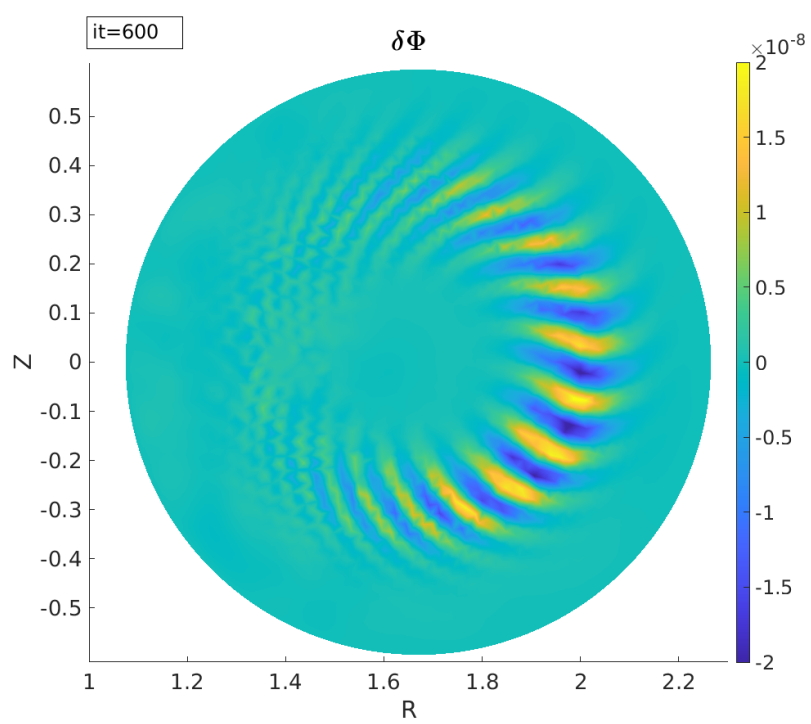}
    \caption{Comparison of the two-dimensional mode structure of the electric field on the poloidal cross section for $n=14$. Left: GPU version. Right: CPU version.}
    \label{2d-mode-structure-comparison}
\end{figure}

To compare the simulation results with and without GPU offloading, tokamak plasma instabilities are simulated. 
In the previous work of TRIMEG-C1 \cite{lu2024gyrokinetic}, the so-called Cyclone case that models the plasma core was used to simulate the ion temperature gradient (ITG) mode in the low $\beta$ limit \cite{gorler2016intercode}. 
The simulation parameters have been adjusted to reduce the computational cost by choosing the normalized ion gyroradius $\rho^*=1/60$, $m_{\rm i}/m_{\rm e}=100$, where $m_{\rm i}$ is the mass of the ion and $m_{\rm e}$ is that of the electron. The gyro average of electrons and ions is switched off. A low $\beta_N=0.004$ is chosen for the plasma to simulate the ITG mode. The simulation uses $8\cdot10^6$ electron markers and $2.5\cdot 10^5$ ion markers to model the plasma behavior.  The number of radial grid points is $N_r=16$ for the toroidal mode number $n=2$ and $N_r=32$ otherwise. We carried out a scan over the toroidal harmonic number, specifically for the values $n=6,10,14,18$. The $n$-scan for the Cyclone case is used to show that the overall implementation of the GPU-accelerated code is correct, including the structural changes that were applied to the code during the GPU porting. The accelerated version was run on Viper GPU, while the original CPU version was run on TOK. Fig. \ref{fig:growth-rate-comparison} shows the agreement between the growth rate $\gamma$ produced by the CPU and GPU versions.  The blue line corresponds to the values of $\gamma$ produced by the CPU version, while the green line identifies those generated by the GPU implementation. The red dotted line corresponds to the percentage difference between each pair of data points. Note that consecutive CPU runs do not produce exactly the same results, so there is an intrinsic statistical variation in the $\gamma$ values. This is due to the initialization of marker quantities by sampling from a Maxwellian distribution using pseudorandom number generators. For this reason, the red line should not be interpreted as the GPU implementation failing to reproduce the correct results, but rather as a metric of the intrinsic randomness in the simulations. It has been assessed that, for the same set of inputs, the GPU kernels produce the exact same results as the CPU computations, when the solver solutions are made to be identical, confirming that there is no race condition within the kernels. Another issue that has been found is a numerical instability affecting the linear systems corresponding to the field equations, which causes the amplification of small round-off errors, potentially compromising the ability to reproduce results with close fidelity. The TRIMEG team is working to diagnose and solve this problem, which does not affect the results of this paper.

The time evolution of the total field energy is shown in Fig. \ref{fig:total-energy-evolution}. The data correspond to the case with toroidal harmonic number $n=14$. Early stages of the evolution show significant differences, but the growth rate closely matches once the exponential phase is reached. Note that the scale of the y-axis is logarithmic; therefore, the trend appears linear. The growth rate of the two versions in the selected time interval, identified by the vertical dotted lines, is reported at the top of the figure, and the difference is $\approx 1.3\%$. 

A comparison of the 2D mode structure is shown in Fig. \ref{2d-mode-structure-comparison}. The tokamak geometry used for this case is simplified and corresponds to an ideal torus. Again, the data correspond to the case with $n=14$, and the overall structure is preserved.

\clearpage
\subsection{Study of the TCV-X21 Case}
%\label{subsec:parameters}
\label{subsec:tcv}
In addition to the Cyclone case, the TCV-X21 case is also studied in this work. The TCV-X21 case has been studied experimentally and numerically for Tokamak à configuration variable (TCV) \cite{oliveira2022validation,body2022development,ulbl2023influence,lu2026PFAFEM,lu2026cpc}. This case is used by various codes such as GRILLIX and GENE-X for the studies of the transport and the profile formation, including the separatrix. In this section, the plasma simulation is performed to demonstrate the features of the nonlinear ITG instability. The electrostatic model is adopted for single-$n$ simulations. The reference Larmor radius is $\rho_{\rm ref}=1$ cm instead of the nominal value $\rho_{\rm ref}=0.3539$ cm, in order to save computational resources. The reference magnetic field is $B_{\rm{ref}}=0.90727$ T. 

\begin{figure}[H]
    \centering
    \includegraphics[width=0.6\linewidth]{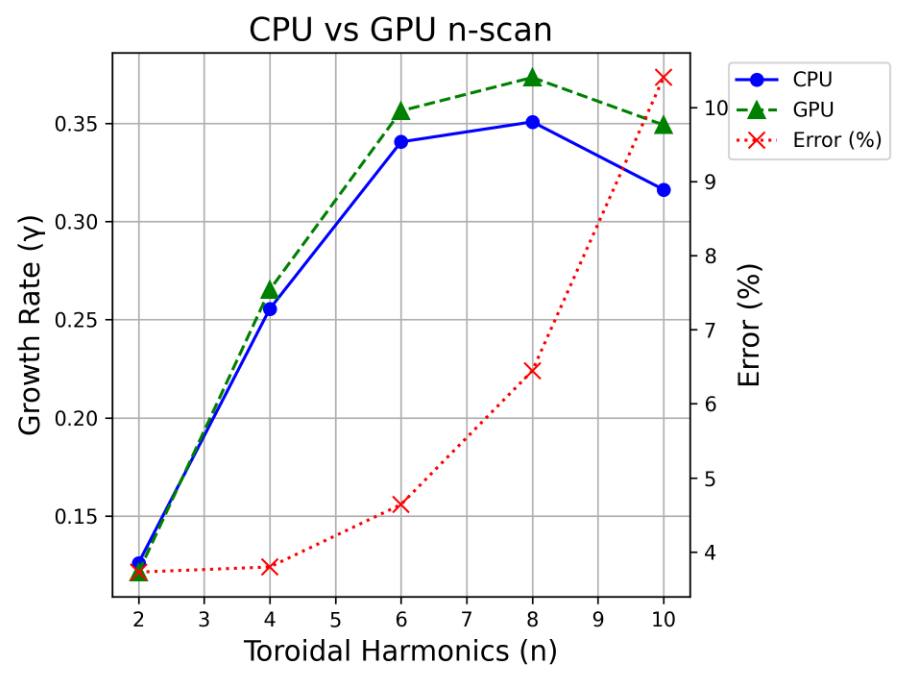}
    \caption{Growth rate comparison between CPU and GPU versions for toroidal mode numbers $n=2, 4, 6, 8, 10$. The error is slightly above 10\% for $n=10$, and below 7\% for the other toroidal mode numbers.}
    \label{fig:tcv-growth-rate}
\end{figure}

\begin{figure}
    \centering
    \includegraphics[width=0.55\linewidth]{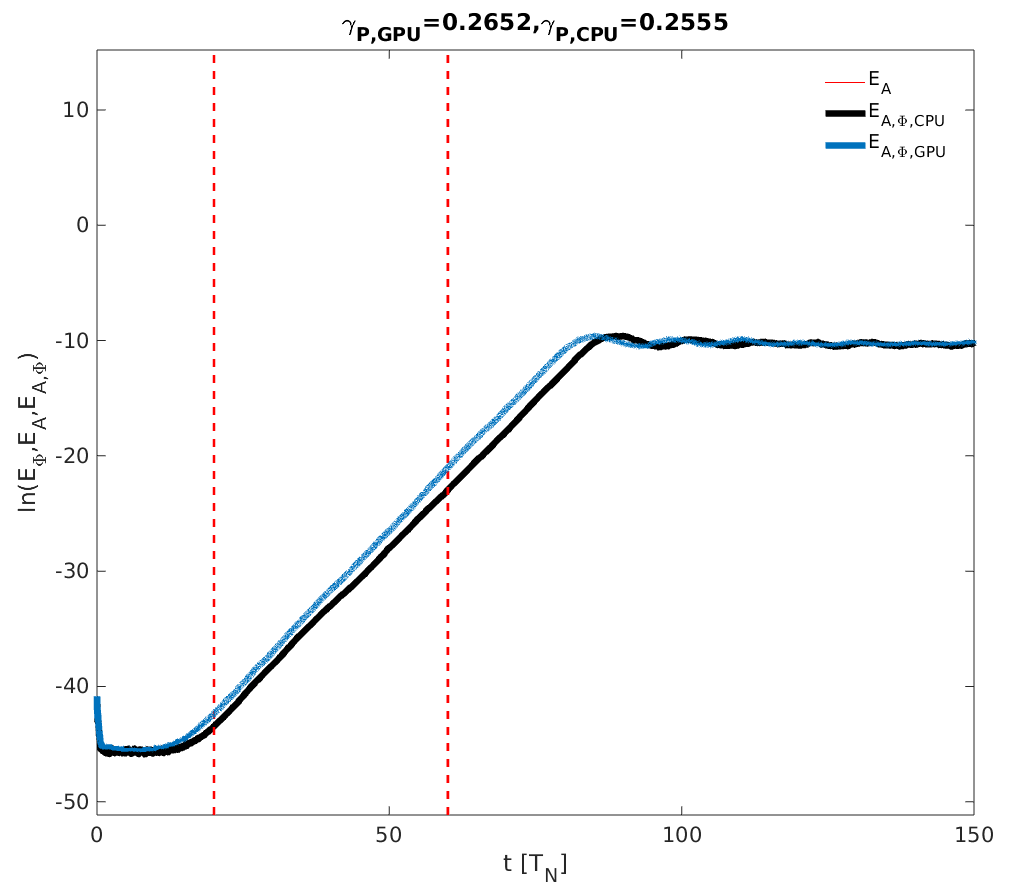}
    \caption{Comparison of the CPU and GPU time evolution of the total energy stored in the electric field for $n=4$. The energy growth rate corresponds to the selected time interval [20, 60] of the linear stage.}
    \label{fig:tcv-energy-evolution}
\end{figure}

\begin{comment}
The corresponding reference density and temperature are $n_{\rm ref}=3.5\times10^{19}/m^3$ and $T_{\rm ref}=600$ keV, respectively, at the reference radial location $\tilde r_0$. \\
\end{comment}
Again, the aim of the $n$-scan is to reproduce the results achieved by the original CPU version with the GPU-accelerated implementation. The former was run on TOK, while the latter was run on the Viper GPU node.
The toroidal harmonics numbers used for the $n$-scan are $2,4,6,8,10$. To keep the simulations economical, the number of electrons is limited to $10^6$, and the number of ions to $2.5\cdot10^5$. This compromise was necessary to complete the $n$-scan in time, but the accuracy is reduced, and the agreement between the results could be improved by increasing these parameters.
The agreement between the growth rates produced by the GPU and CPU versions is shown in Fig. \ref{fig:tcv-growth-rate}. As before, the blue line corresponds to the values of $\gamma$ produced by the CPU version, while the green line identifies the values generated by the GPU implementation. The red dotted line plots the percentage difference between each pair of data points. The agreement deteriorates for higher values of $n$, but the variation of the growth rate as a function of the toroidal mode number is correctly reproduced by the GPU implementation.\\
The time evolution of the electric field from the exponentially growing stage to the nonlinear saturation stage is shown in Fig. \ref{fig:tcv-energy-evolution}. The data correspond to the case with the toroidal harmonic number $n=4$. The linear growth rate shows good agreement, and the saturation level of the nonlinear stage is the same for both simulations, indicating eventual convergence to the same value even if the previous stages are not identical. The two vertical dotted lines identify the time interval used to compute the linear growth rate, which is reported at the top of the plot.

\begin{figure}[h!]
    \centering
        \includegraphics[width=.35\linewidth]{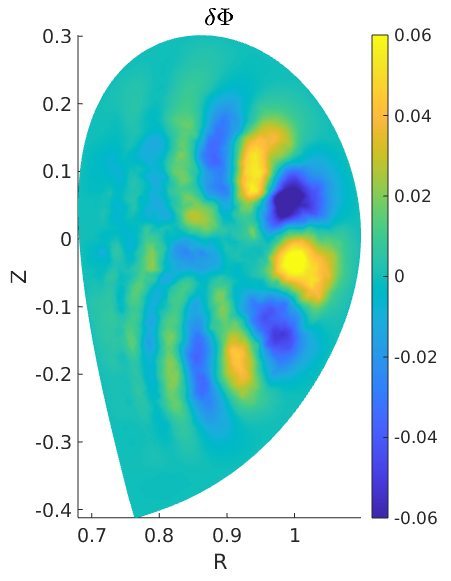}
        \includegraphics[width=.35\linewidth]{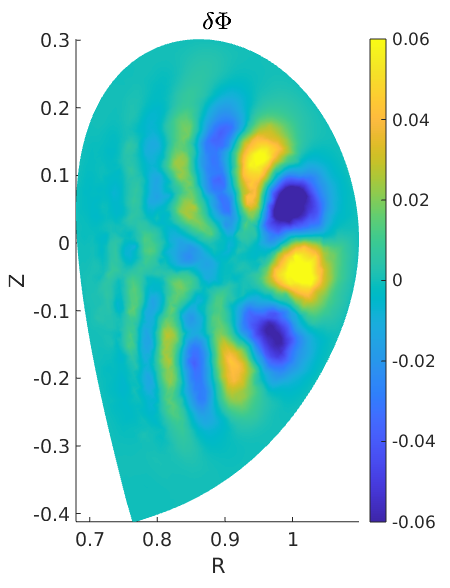}
        \caption{Comparison of the two-dimensional mode structure of the electric field on the poloidal cross section for $n=4$. Left: GPU version. Right: CPU version.}
    \label{tcv-mode-structure}
\end{figure}

Finally, a comparison of the 2D mode structure in the poloidal cross section is shown in Fig. \ref{tcv-mode-structure}. The tokamak geometry is more realistic than that used in the Cyclone case, and the simulation's outer boundary coincides with the separatrix. Again, the data correspond to the case with $n=4$. The GPU version reproduces the overall mode structure correctly, even though there are minor differences.

\ifpdf
    \graphicspath{{Chapter4/Figures/PNG/}{Chapter3/Figures/PDF/}{Chapter4/Figures/}}
\else
    \graphicspath{{Chapter4/Figures/EPS/}{Chapter3/Figures/}}
\fi

\newpage
\section{\label{sec:conclusions} Conclusions and future work}

In this work, a computationally expensive portion of the TRIMEG-C1 gyrokinetic PIC code has been successfully ported to GPU architectures using the OpenMP offloading framework, achieving a portable implementation on both NVIDIA and AMD platforms. The particle push kernel and the subroutines performing grid-to-particle (G2P) operations were identified as the primary candidates for GPU acceleration, due to their inherent parallelism over independent marker particles and their dominant contribution to the total simulation time in large-scale runs. 
The OpenMP API was selected as the offloading paradigm because it is currently the only practical approach for supporting both NVIDIA and AMD GPU architectures in Fortran while minimizing divergence from the original code.

The porting process required extensive code restructuring to accommodate the limitations of the \texttt{nvfortran} and \texttt{amdflang} compilers, neither of which fully supports advanced Fortran features inside OpenMP GPU kernels.

The performance of the GPU implementation was evaluated through a comprehensive set of benchmarks on multiple clusters. For a realistic problem size of $32 \times 10^6$ marker particles on a full node, the particle push kernel achieves speedups of $\approx9$ on both Viper and Pitagora, based on the AMD MI300A APU and NVIDIA H100 GPUs respectively, compared to the fastest CPU-only implementation. The strong scaling analysis demonstrated that the GPU-accelerated code scales well across multiple nodes.

The physical correctness of the GPU implementation was verified against two well-established benchmark cases. For the Cyclone base case \cite{gorler2016intercode}, which simulates the ITG mode in a simplified circular tokamak geometry, a scan over toroidal mode numbers $n = 6, 10, 14, 18$ showed good agreement between the energy growth rates and two-dimensional mode structures obtained with the CPU and GPU versions. For the TCV-X21 case \cite{oliveira2022validation}, which simulates nonlinear electrostatic ITG instabilities in a realistic diverted geometry including the separatrix, a scan over $n = 2, 4, 6, 8, 10$ toroidal harmonic numbers confirmed that the GPU version reproduces the linear growth rates and nonlinear saturation levels with good fidelity.

Future work will focus on several directions. First, the integration of GPU-enabled solvers through \texttt{PETSc} will also be considered, using backends such as CUDA or HIP. Second, a C++ port of TRIMEG will be considered to improve GPU offloading and CPU-only performance. Initial OpenMP offloading results for the C++ version are promising, achieving better performance with substantially lower porting effort. Additionally, GPU offloading of the particle-to-grid (P2G) operations is being implemented in the C++ code.
The GPU-accelerated TRIMEG-C1 provides the computational foundation for future large-scale physics studies, including electromagnetic simulations at realistic values of $\rho^*$ and mass ratio, multi-scale turbulence studies, and investigations of plasma transport in diverted tokamak configurations.

%%%%%%%%%%%%%%%%%%%%%%%%%%%%%%%%%%%%%%%%%%%%%%%%%%%%%%%%%%%%%%%%%%
\ack{
This work has been carried out within the framework of the EUROfusion Consortium, funded by the European Union via the Euratom Research and Training Programme (Grant Agreement No 101052200—EUROfusion). Views and opinions expressed are however those of the author(s) only and do not necessarily reflect those of the European Union or the European Commission. Neither the European Union nor the European Commission can be held responsible for them.
}

%%\clearpage
%\null

%\appendix
%\section{}
%\label{}

%\funding{Sample text inserted for demonstration.}
% This section is a list of funder names and grant numbers

%\roles{Sample text inserted for demonstration.}
% List author names and the contributions made to the article, using terms from the NISO Contributor Roles Taxonomy (CRediT) https://credit.niso.org

%\data{Sample text inserted for demonstration.}
% For more information on IOP Publishing's research data policy see: https://publishingsupport.iopscience.iop.org/questions/research-data/

%\suppdata{Sample text inserted for demonstration.}

%\section*{References}

%\bibliographystyle{elsarticle-num-names} 
\bibliographystyle{iopart-num}
\bibliography{reference}

\end{document}